\documentclass[
    ,final            
  ]
  {aipproc}

\layoutstyle{6x9}

\usepackage{graphicx}
\usepackage{epsfig}
\usepackage{amsbsy}
\usepackage[rightcaption]{sidecap}

\newcommand{\proton}{\ensuremath{\mathrm{p}}}

\newcommand{\pip}{\ensuremath{\pi^{+}}}
\newcommand{\pim}{\ensuremath{\pi^{-}}}
\newcommand{\pizero}{\ensuremath{\pi^{0}}}
\newcommand{\kp}{\ensuremath{K^{+}}}

\newcommand{\SigmaPlus}{\ensuremath{\Sigma^{+}}}
\newcommand{\SigmaZero}{\ensuremath{\Sigma^{0}}}
\newcommand{\SigmaMinus}{\ensuremath{\Sigma^{-}}}
\newcommand{\LambdaOne}{\ensuremath{\Lambda(1405)}}

\newcommand{\LambdaTwo}{\ensuremath{\Lambda(1520)}}

\newcommand{\Kstar}{\ensuremath{K^{\ast}}}

\renewcommand{\d}{\ensuremath{\mathrm{d}}}

\newcommand{\etal}{\textit{et al.}}

\begin{document}

\title{Properties of the
  $\boldmath{\Lambda}$($\mathbf{1405}$) Measured
  at CLAS}


\author{Kei Moriya}{
  address={Department of Physics, Indiana University,
    Bloomington, IN 47405-7105}
}

\author{Reinhard Schumacher}{
  address={Department of Physics, Carnegie Mellon University,
    Pittsburgh, PA 15213 \\
  (for the CLAS Collaboration)}
}

\begin{abstract}
The nature of the \LambdaOne, and its place in the baryon spectrum has
remained uncertain for decades. Theoretical studies have shown that it
may possess strong dynamical components which are not seen in other
well-known baryons. Using the CLAS detector system in Hall B at
Jefferson Lab, we have measured the photoproduction reaction
$\gamma + \proton \to \kp \Lambda(1405)$ with high statistics and over
different $\Sigma \pi$ decay channels. The reconstructed invariant
mass distribution (lineshape) has been measured, as well as the
differential cross sections for the \LambdaOne, $\Sigma(1385)$, and
\LambdaTwo. Our analysis method is discussed and our near-final
results for the \LambdaOne{} lineshape and differential cross section
are presented.
\end{abstract}

\maketitle



The spectrum of baryons has in the past
played an important role in the establishment of the quark
model. Today it is known that this spectrum matches very well to the
so-called constituent quark model at low excitation energies.
The \LambdaOne{} is unique in the sense that although it is only the
first excited $\Lambda$ state and has relatively low excitation
energy, the nature of this resonance is not well established. Past
experiments have shown a distortion of the
invariant mass distribution (``lineshape'') of the \LambdaOne{} from a
simple Breit-Wigner form. On the theoretical side,
all analyses agree that the closeness of the nearby $N \overline{K}$
threshold plays a role in the distortion, there is no universally
accepted explanation (see~\cite{Dalitz:PDG} for a review). For the
constituent quark model, the \LambdaOne{} has the largest discrepancy
between the model prediction and experimental
mass~\cite{Isgur-Karl_PRD18}.

Until recently, there was not enough experimental data on the
\LambdaOne{} to gain further insight into its nature. Now, using the
CLAS~\cite{CLAS-NIM} detector system in Hall B at the Thomas Jefferson
National Accelerator Facility, we have done a photoproduction
measurement of the \LambdaOne{} with unprecedented statistics. With
the high statistics and good resolution of the CLAS system, we
were able to extract the \LambdaOne{} lineshapes and differential
cross sections in all three of its $\Sigma \pi$ decay channels. Our
measurements give a much improved quantitative description of the
\LambdaOne, and may lead to further insights into its nature.

\section{Theory of the $\boldmath{\Lambda}$($\mathbf{1405}$)}

\begin{SCfigure}
  \centering
  \includegraphics[width=.47\textwidth,angle=90]{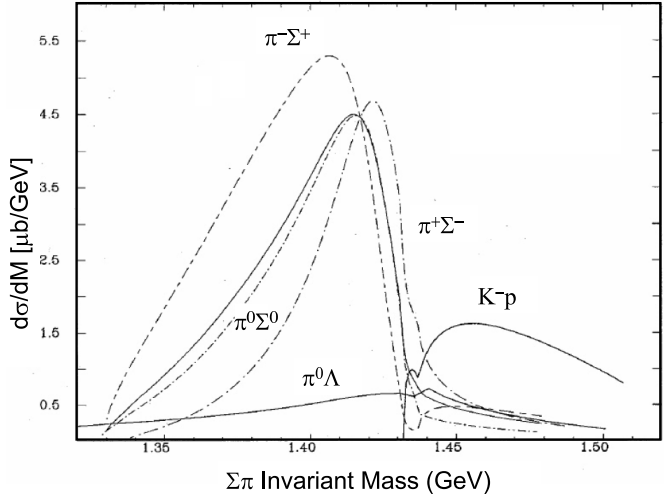}
  \caption{Theoretical lineshapes predicted~\cite{Nacher} for the
    \LambdaOne{} using one particular chiral unitary approach. Due to
    the
    interfering isospin $0$ and $1$ terms, the lineshapes are
    distorted from a simple Breit-Wigner form. In more recent years,
    attention has also focused on the possible two pole structure of
    the \LambdaOne{} in the $I=0$ channel alone. This too is expected
    to influence the lineshapes~\cite{Oller-Meissner,Jido}.}
  \label{fig:theorylineshape}
\end{SCfigure}

In recent years, there has been a renewed interest in the \LambdaOne,
especially from the point of view of chiral
dynamics~\cite{Kaiser-Siegel-Weise,Oset-Ramos,Oller-Meissner}, where
the \LambdaOne{} is dynamically generated. More recent developments
show that there may be two poles for the
\LambdaOne~\cite{Oller-Meissner,Jido}, which may lead to
different lineshapes of the \LambdaOne{} for different reactions. A
paper by Nacher \etal~\cite{Nacher} has predicted the
lineshapes of the \LambdaOne{} for photoproduction near threshold,
which is shown in Figure~\ref{fig:theorylineshape}. The
prediction is that not only are the lineshapes of the \LambdaOne{}
distorted from a simple Breit-Wigner form, but different decay
channels have different shapes due to the coupled channels mixing
isospin contributions. Our main goal for this analysis has been on
comparing the experimental data with this prediction.

\section{Data Analysis Using CLAS and Results}

Using the CLAS detector at Jefferson Lab, we have obtained a large
dataset containing \LambdaOne{} events by tagging real photons from
$\kp \LambdaOne$ threshold up to $2.84$ GeV in center-of-mass
energy. Charged particles decaying from the reaction $\gamma + \proton
\to \kp \LambdaOne$, $\LambdaOne \to \Sigma \pi$ were detected for the
decay channels $\SigmaPlus \pim$, $\SigmaZero \pizero$, and $\SigmaMinus
\pip$.

In this analysis the main background reactions are $\gamma + \proton
\to \kp \Sigma(1385)$, where the $\Sigma(1385)$ overlaps significantly
with the \LambdaOne{} in the $\Sigma \pi$ invariant mass spectrum, and
the reaction $\gamma + \proton \to \Kstar \Sigma$, where the \Kstar{}
can have kinematic overlap with the \LambdaOne. The $\Sigma \pi$
invariant mass spectrum was
individually fit in bins of center-of-mass energy and center-of-mass
kaon angles to Monte Carlo templates of the reactions of interest. The
contributions thus determined for the $\Sigma(1385)$, \Kstar, and
\LambdaTwo{} were subtracted to obtain the yield for the \LambdaOne,
which was then acceptance-corrected.

A recent development in our analysis has been a final iteration of our
Monte Carlo templates used for the \LambdaOne, where the results of
our lineshapes determined from the data were used as direct input for
generating our Monte Carlo events. This has helped in improving our
fit quality for the $\Sigma \pi$ spectra, and we have confirmed that
the templates used for the \LambdaOne{} have little influence on our
results. Another development includes taking into account the decay of
ground state hyperons that decay outside of the CLAS Start Counter, which
was part of our trigger. This mainly affects data for the
$\Sigma(1385)$ in its decay to $\Lambda \pizero$, and this raises the
measured differential cross section of the $\Sigma(1385)$ by $\sim
3\%$.

\begin{figure}[h!t!b!p!]
  \includegraphics[width=\textwidth]{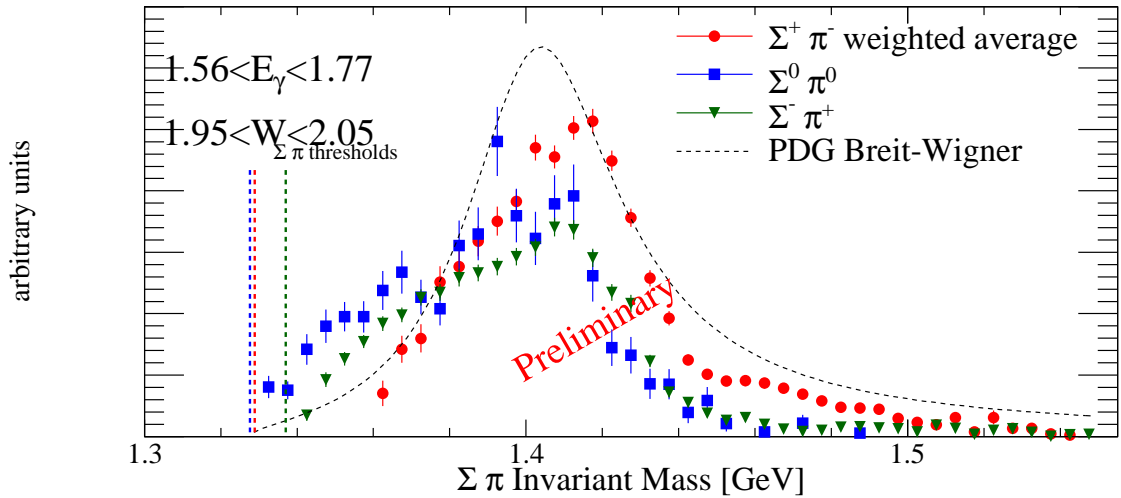}
  \caption{(Color online) The $\Sigma \pi$ invariant mass spectrum measured for the
    \LambdaOne{} for the energy range $1.95<W<2.05$ GeV. The different
    decay channels are shown as $\SigmaPlus
    \pim$ (red circles), $\SigmaZero \pizero$ (blue squares), and
    $\SigmaMinus \pip$ (green triangles). An example of a relativistic
    Breit-Wigner function using the PDG~\cite{PDG2011} values for the
    \LambdaOne{} mass and width are shown as the dashed line.}
  \label{fig:dsigmadM_W1}
\end{figure}

Our results for the lineshape of the \LambdaOne{} are summed over all
center-of-mass kaon angles for each $100$ MeV-wide energy bin we have
in center-of-mass energy. Figure~\ref{fig:dsigmadM_W1} shows the
lineshape of the \LambdaOne{} for each of the $\Sigma \pi$ decay
channels that we have measured for $1.95<W<2.05$. Note that in this
energy bin close to the $\kp \LambdaOne$ threshold, we are just below
the edge of the nominal \Kstar{} threshold, and the \Kstar{} has very
limited kinematic influence. The lineshapes are seen to be quite
different from each other, with the $\SigmaPlus \pim$ channel peaking
at a mass of $\sim 1420$ MeV, and having a much more narrow structure
than the $\SigmaMinus \pip$ channel. This is in contrast to the
theoretical prediction of Figure~\ref{fig:theorylineshape}, where the
$\SigmaMinus \pip$ channel is expected to peak at a higher mass and
have a narrow structure compared to the $\SigmaPlus \pim$ channel.

\begin{figure}[h!t!b!p!]
  \centering
  \includegraphics[width=\textwidth]{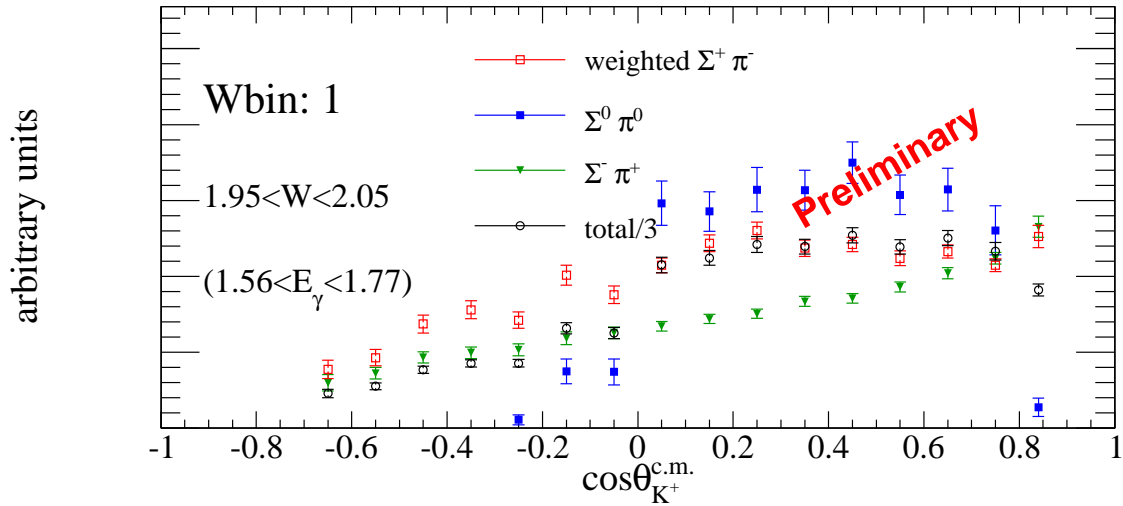}
  \caption{(Color online) The differential cross section $\d \sigma /
    \d \cos \theta_{\kp}^{\mathrm{c.m.}}$ of the
    \LambdaOne{} for the decay channels $\SigmaPlus \pim$ (red empty
    squares), $\SigmaZero \pizero$ (blue filled squares), and
    $\SigmaMinus \pip$ (green triangles). The average of the three
    channels is shown as the black empty circles.}
  \label{fig:xsec1405_W1}
\end{figure}

By summing the lineshape over the $\Sigma \pi$ invariant mass range,
we are able to calculate the differential cross section of the
\LambdaOne. This is shown in Figure~\ref{fig:xsec1405_W1} for the same
energy bin as Figure~\ref{fig:dsigmadM_W1}. The
differential cross sections for each channel are shown in various
symbols, and the behavior of the $\SigmaPlus \pim$ and $\SigmaMinus
\pip$ channels are again seen to be different from each other. This
complicated behavior of the differential cross section depending on
decay channel may indicate that the strong dynamics that create the
\LambdaOne{} has dependence on the center-of-mass angle at which it is
produced.

\section{Conclusion}

The photoproduction differential cross section of the \LambdaOne{} has
been measured using the CLAS detector system at Jefferson Lab's Hall
B. Our lineshape results show a strong difference between each $\Sigma
\pi$ decay channel, which has been predicted in the chiral unitary
coupled channel approach, albeit with channels being interchanged from
the prediction. The \LambdaOne{} remains an exciting research topic,
as recently there have also been studies of seeing how the lineshape
can change depending on momentum transfer $Q^{2}$ in
electroproduction~\cite{Haiyun}. The current photoproduction analysis
is in the stages of collaboration review, and we hope to finalize
and present our final results in the near future. Our data may help
in shedding light on the nature of the \LambdaOne, and its production
mechanism.


\begin{theacknowledgments}
  K.~M. would like to thank the organizers of the NSTAR11 conference
  for the opportunity to present results of our analysis, and also for
  financial support during the conference.
\end{theacknowledgments}



\bibliographystyle{aipproc}   

\bibliography{mybiblio}

\IfFileExists{\jobname.bbl}{}
 {\typeout{}
  \typeout{******************************************}
  \typeout{** Please run "bibtex \jobname" to optain}
  \typeout{** the bibliography and then re-run LaTeX}
  \typeout{** twice to fix the references!}
  \typeout{******************************************}
  \typeout{}
 }

\end{document}